\documentclass[review]{elsarticle}

\usepackage{hyperref}

\usepackage[symbol]{footmisc}
\usepackage{float}
\usepackage{caption}
\usepackage{subcaption}

\journal{Journal of \LaTeX\ Templates}

\begin{document}

\begin{frontmatter}

\title{Comparing X-ray color selection in separating X-ray binary classes using Color-Color-Intensity diagrams}

\author{N.Islam$^1$\footnote[2]{Present affiliation: 1. Center for Space Science and Technology, University of Maryland,
Baltimore County, 1000 Hilltop Circle, Baltimore, MD 21250, USA \\
2. X-ray Astrophysics Laboratory, NASA Goddard Space Flight Center,
Greenbelt, MD 20771, USA}, S.D. Vrtilek$^1$\footnote[1]{Corresponding authors E-mail:  nislam@umbc.edu (N. Islam); svrtilek@cfa.harvard.edu (S.D. Vrtilek)}, Bram Boroson$^2$, D.-W. Kim$^1$, E. O'Sullivan$^1$, M. L. McCollough$^1$, G. Fabbiano$^1$, C. Anderson$^1$, D. J. Burke$^1$, R. D'Abrusco$^1$, A. Fruscione$^1$, J. L. Lauer$^1$, D. Morgan$^1$, A. Mossman$^1$, A. Paggi$^{3,4,5}$, G. Trinchieri$^6$}

\address{$^{1}$Center for Astrophysics $\mid$ Harvard $\&$ Smithsonian, 60 Garden Street, Cambridge, MA 02138, USA\\
$^{2}$Department of Physics and Chemistry, Clayton State University, GA, USA\\
$^{3}$Dipartimento di Fisica, Universit\`{a} degli Studi di Torino, via Pietro Giuria 1, I-10125 Torino, Italy\\
$^{4}$Istituto Nazionale di Fisica Nucleare, Sezione di Torino, via Pietro Giuria 1, I-10125 Torino, Italy\\
$^{5}$INAF- Osservatorio Astrofisico di Torino, via Osservatorio 20, 10025 Pino Torinese, Italy\\
$^{6}$INAF-Osservatorio Astronomico di Brera, via Brera 28, 20121 Milano, Italy}

\begin{abstract}
X-ray binaries exhibit a wide range of properties but there are few accepted methods to determine the nature of the compact object. Color-Color-Intensity diagrams have been suggested as a means of distinguishing between systems containing black holes from those containing neutron stars. However, this technique has been verified with data from only one instrument (RXTE/ASM) with a single set of X-ray colors defined using data available only in pre-determined energy bands. We test a selection of X-ray colors with a more sensitive instrument to determine the reliability of this method.
We use data from the {\it MAXI}/Gas Slit Camera, which allows users to specify energy-bands. We test X-ray colors that have been previously defined in the literature as well as ones that we define specifically in this paper. 
A representative set of systems are used to construct Color-Color-Intensity diagrams in each set of colors to determine which are best for separating different classes. For studying individual sources certain bands are more effective than others. For a specified energy range, the separation of soft states in black hole binaries was possible only where both soft and hard colors included information from the lowest energy band. We confirm that Color-Color-Intensity diagrams can distinguish between systems containing black holes or neutron stars in all X-ray colors tested; this suggests an universality in the accretion processes governing these different classes. We suggest possible physical processes driving different classes of X-ray binaries to different locations in Color-Color-Intensity diagrams.
\end{abstract}

\begin{keyword}
methods: data analysis -- X-rays: binaries -- stars: neutron -- stars: black holes
\end{keyword}

\end{frontmatter}



\section{Introduction}

 X-ray binaries (XRBs) are gravitationally bound binary systems, consisting of a compact object  together with a
main sequence or supergiant companion. The compact object can be either a black hole (BH) or a neutron star (NS).  Accretion, in which matter from the companion is captured by the compact 
object either from a stellar wind or through Roche lobe overflow, is the main source of energy for these systems.
XRBs are often identified by the mass of the companion star: when the mass of the companion star is $\geq$ 10 M$_{\odot}$ they are called High Mass X-ray Binaries (HMXBs); when the mass of the companion is $\leq$ 1 M$_{\odot}$ they are called 
Low Mass X-ray Binaries (LMXBs). NS systems that are highly magnetized show X-ray pulses and are referred to as X-ray pulsars (for review see \citealt{nagase1989,paul2017}). Both HMXBs and LMXBs exhibit a wide range of temporal and spectral properties
which depend on the nature of the compact object, the type of the companion star, the nature of the accretion process, 
magnetic field strengths, etc.
\par
X-ray Color-Color (CC) and Color-Intensity (CI) diagrams are widely used to
study XRBs. X-ray colors are defined as the ratio of counts in two X-ray energy bands from the instrument used. CC diagrams represent spectral variations of XRBs, and CI
diagrams represent their intensity variations over spectral states. Patterns traced
by XRBs in these plots serve as diagnostic tools in classifying classes or spectral
states. For example, non-pulsing (weakly magnetised) NS LMXBs are classified
as Z or Atoll sources based on the patterns they form in CC plots \citep{hasinger1989}.   \cite{homan2010} noted that a single source can show both Z or Atoll patterns. \cite{homan2010} and \cite{fridriksson2015} suggest that the primary physical process determining the patterns is mass accretion rate. Black hole (BH) binaries with both high-mass (HMBH) and low-mass (LMBH) companions, have two main spectral states: a soft state, dominated by thermal emission from an accretion disc, and a hard state, with Comptonised spectra; many show a third state intermediate between the other two \citep{remillard2006}. For LMBHs such as GX 339--4, the different X-ray spectral states often trace a “q” shaped track in CI diagrams \citep{fender2004}. Differences between the behavior of a source on different tracks of a CC or CI plot can also be manifested in detailed timing and spectral studies. For example, \cite{munoz2014}  suggest that source types can be distinguished by their location in fast variability-luminosity diagrams.
\par
\cite{white1984} plotted all X-ray binaries observed by the A2 instrument
on the High Energy Astrophysical Observatory \citep{rothschild1979} on a CC
plot and found that one extreme of the plot showed only systems containing BHs
and another extreme showed only X-ray pulsars, but in the center of their plot,
systems containing BHs, pulsars, and non-pulsing NSs all overlapped.  \cite{done2003} (hereafter DG03) fitted models to segments of CC and CI diagrams of bright sources in order to determine the physical behavior that put sources in a given location; they showed that certain regions that are occupied by black holes are “inaccessible” to neutron stars and attributed this to the presence of boundary layer emission in neutron stars.
\par
Vrtilek $\&$ Boroson (2013; hereafter VB13) showed that combining CC and CI diagrams in a single three-dimensional Color-Color-Intensity (CCI) plot distinguishes key structures that can be obscured in the two-dimensional projections enforced by CC or CI plots.
In a CCI diagram, the various classes of X-ray binaries -- systems containing white dwarfs, neutron stars, or black holes -- separate into complex, but geometrically distinct volumes. Using CCI diagrams, \cite{gopalan2015} developed a probabilistic (Bayesian) model which uses a supervised learning approach (unknown classifications are predicted using known classifications as priors), to predict the type of an unknown X-ray binary. 
\par
VB13 suggest that CCI diagrams are three-dimensional counterparts for X-ray binaries of the classic Hertszprung-Russell (HR) diagram, which separates different types of single stars and proved fundamental to our understanding of single star evolution. 
However, the definition of ``X-ray color'' varies from instrument to instrument and different authors sometimes define different ratios for a given instrument (e.g \citealt{peris2016,koljonen2010,homan2010,fender2004,belloni2000}).
Variations are to some extent due to the different ranges available in different instruments, {\it e.g}, the {\it RXTE}/ASM (\citealt{levine1996}) light-curves are provided by the instrument's team in three pre-defined energy bands: 1.5--3 keV, 3--5 keV and 5--12 keV; the X-ray telescope {\it Chandra} is most sensitive in the energy band of 0.3--8 keV. 
Analogous to the HR diagram, investigating the physical reasons why different classes of X-ray binaries occupy different locations in CCI diagrams, as done by DG03 for CC and CI diagrams, would be a significant step in understanding accreting binaries.
\par
VB13 had data available only in pre-determined energy bands as provided by the RXTE/ASM team defining a single set of X-ray colors.  Here we use data from the Gas Slit Camera (GSC; \citealt{mihara2011}) on board the Monitor of All sky X-ray Image (MAXI; \citealt{matsuoka2009}) which has the highest sensitivity and energy resolution among past and currently active all sky monitors. The GSC also provides the ability for data to be extracted in user-specified energy bands.
\par
In this paper, we extract data from the MAXI/GSC in a range of energy bands defining 10 different sets of X-ray colors. We construct CCI diagrams for a subset of XRBs representative of the major classes. In Section 2 we present the observations and method of analysis.  In Section 3 we compare the effects of different X-ray colors on the separation of XRB classes using CCI diagrams. In Section 4 we compare the ability of different X-ray colors to separate spectral states observed in BH XRBs using CCI diagrams.  In Section 5 we suggest possible physics driving a system to a given location in CCI diagrams.  We discuss these results and present our conclusions in Section 6.


\section{Data and Analysis}

\subsection{Observations}
\label{sec:obs} 

{\it MAXI} has been operating on the Kibo module of the International Space Station since August 2009. 
The GSC, the main instrument on board MAXI, operates in the energy range 2--30 keV. It consists of six units of large area position sensitive Xenon proportional counters. The in-orbit performance of {\it MAXI}/GSC is summarised in \citealt{sugizaki2011}. The GSC reaches a 1 day sensitivity of 9 mCrab (3 $\sigma$), compared to 15 mCrab for {\it RXTE}/ASM (\citealt{levine1996}) and 16 mCrab for {\it Swift}/BAT \citep{krimm2013}. 
The GSC typically scans a point source on the sky during a transit of 40 -- 150 seconds during its 92 minutes orbital period. 
\par
 We have extracted one-day averaged count-rates, having at least 3$\sigma$ significance in each energy-band, for each of the sources listed in Table 1, in the X-ray colors defined in Table 2, using the MAXI on-demand processing\footnote{http://maxi.riken.jp/mxondem/}. The sample of bright XRBs in Table 1 have been observed and extensively studied with several X-ray observatories and represent the major ``classes" of XRBs:  BH systems with high and low mass companions (HMBHs and LMBHs), pulsing NS systems with high and low mass companions (HMXB and LMXB pulsars), and non-pulsing NS systems (Z and Atoll LMXBs). Where possible we have included four sources in each category. However, we have only three confirmed HMBHs (Cyg X--1, LMC X--1 and LMC X--3) and only three LMXB pulsars with sufficient statistics in MAXI/GSC (4U 1626--67, 4U 1822--37 and Her X--1).
\par
\begin{figure}[htb]
    \begin{minipage}[t]{.45\textwidth}
        \centering
        \includegraphics[scale=0.3,angle=0]{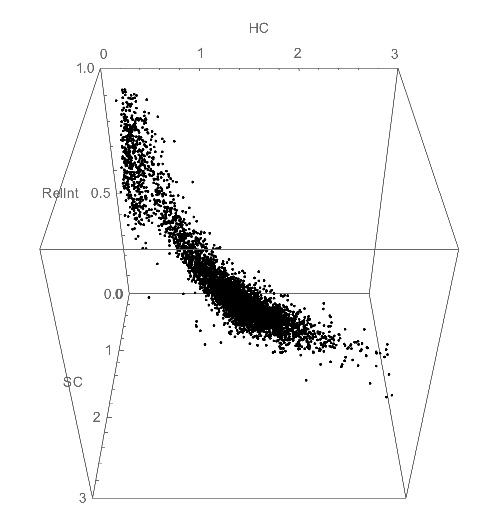}
        \subcaption{}
    \end{minipage}
    \hfill
    \begin{minipage}[t]{.45\textwidth}
        \centering
        \includegraphics[scale=0.3,angle=0]{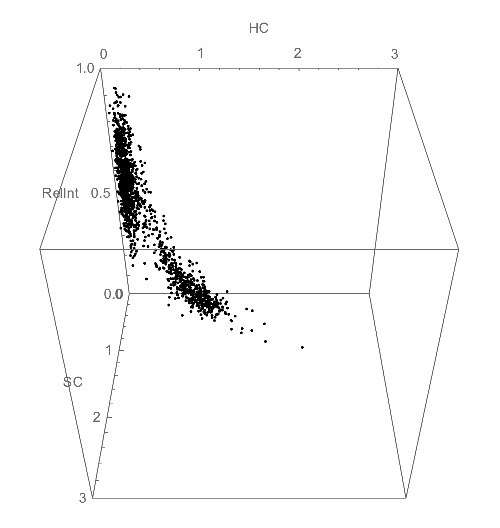}
        \subcaption{}
    \end{minipage}  
    \caption{Color-Color-Intensity diagrams of Cyg X--1 {\bf (a):} Using RXTE/ASM data in the bands provided by the ASM Team. 13 years of data are plotted as one day averages.  {\bf (b):} Using MAXI/GSC data extracted in colors closest to that of the ASM (CR1). 8 years of data are plotted as one day averages. The coordinates SC and HC are as defined in Table 2; the coordinate RelInt is the sum of the counts in all energy bands divided by the average of the top 1\% of the counts for each source. The CCI plots show consistent patterns in both instruments.}
    \label{compare_asm}
\end{figure}

The X-ray colors used in this paper are listed in Table 2. Since MAXI has a low-energy cutoff of 2 keV, we cannot reproduce the exact energy range of the RXTE/ASM (1.3-12 keV). Instead we directly compare actual ASM data with MAXI data extracted in bands as close to the ASM as possible (CR1 in Table 2). 
Figure \ref{compare_asm} shows plots of Cyg X--1 (a) using 13 years of actual RXTE/ASM data in the energy bands provided and (b) using 8 years of MAXI/GSC data in the energy bands closest to the ASM bands (CR1).  This demonstrates that the bands defined for MAXI correlates very well with that provided by the ASM. CR2 is similar to CR1, except that the highest energy band is at 8 keV (to match limits for {\it Chandra}, which will be used for a future work mentioned in Section 6), CR3 and CR4 have been previously defined using {\it RXTE}/PCA in \cite{homan2010}, and \cite{peris2016} respectively.  Colors CR5, CR6, CR7, CR8, CR9, and CR10 are defined in this paper to test additional color combinations. X-ray colors CR1 -- CR4 are used in Section 3 where we compare CCI diagrams of various classes of XRBs.  X-ray colors CR5 -- CR10 are used in Section 4 where we test the effects of different colors on separation of spectral states of BH binaries.

\begin{table*}
\centering
\caption{X-ray Binary Sources}
\label{XRBs}
\begin{tabular}{lccc}
\hline
Source&Compact&Companion&Source\\
Name&Object Type&Mass&Class\\
\hline
\hline
Cyg X--1 & Black Hole & High & HMBH\\
LMC X--1 & Black Hole & High&HMBH\\
LMC X--3 & Black Hole & High & HMBH\\
\hline
GX 339--4&Black Hole&Low& LMBH\\
GRS1915+105&Black Hole&Low& LMBH\\
H1743--32&Black Hole&Low&LMBH\\
4U 1630--472&Black Hole&Low&LMBH\\
\hline
Cyg X--2&Neutron Star&Low&Z source\\
GX 17+2&Neutron Star&Low&Z source\\
GX 5--1 &Neutron Star&Low&Z source\\
Sco X--1&Neutron Star&Low&Z source\\
\hline
GX9+9&Neutron Star&Low&Atoll source\\
GX9+1&Neutron Star&Low&Atoll source\\
Aql X--1&Neutron Star&Low&Atoll source\\
4U 1608--52&Neutron Star&Low&Atoll source\\
\hline
Cen X--3&Neutron Star&High&HMXB pulsar\\
Vela X--1&Neutron Star&High&HMXB pulsar\\
SMC X--1&Neutron Star&High&HMXB pulsar\\
4U 1538--52&Neutron Star&High&HMXB pulsar\\
\hline 
4U 1626--67&Neutron Star&Low&LMXB pulsar\\
4U 1822--37&Neutron Star&Low&LMXB pulsar\\ 
Her X--1&Neutron Star&Low&LMXB pulsar\\
\hline
\end{tabular}
\end{table*}

\begin{table*}
\centering
\caption{X-ray color definitions} 
\label{colors}
\begin{tabular}{lcccc}
\hline
Label & Energy bands used (keV) & soft color (SC) & hard color (HC) \\
\hline
CR1& 2-3;3-5;5-12 & (3-5)/(2-3) & (5-12)/(2-3)\\
CR2& 2-3; 3-5; 5-8 & (3-5)/(2-3) & (5-8)/(2-3)\\
CR3& 2.4-4; 4-7.3; 7.3-9.8; 9.8-18.2   &   (4-7.3)/(2.4-4) & (9.8-18.2)/(7.3-9.8)\\
CR4& 2.2-3.6; 3.6-5; 5-8.6; 8.6-18   &  (3.6-5)/(2.2-3.6) & (8.6-18)/(5-8.6)\\
CR5& 2-7; 7-12; 12-18 & (7-12)/(2-7) & (12-18)/(2-7)\\
CR6& 2-7; 7-12; 12-15; 15-18 & (7-12)/(2-7) & (15-18)/(12-15)\\
CR7& 2-5; 5-8; 8-18  &  (5-8)/(2-5) & (8-18)/(2-5)\\
CR8& 2-5; 5-8; 8-13; 13-18 & (5-8)/(2-5) & (13-18)/(8-13)\\
CR9& 2.4-5; 5-8; 8-18.2 & (5-8)/(2.4-5) & (8-18.2)/(2.4-5) \\
CR10& 2.2-5; 5-8; 8-18 &(5-8)/(2.2-5)&(8-18/2.2-5)\\
\hline
\end{tabular}

\end{table*}

\begin{figure}[htb]
    \begin{minipage}[t]{.2\textwidth}
        \centering
       \includegraphics[scale=0.18,angle=0]{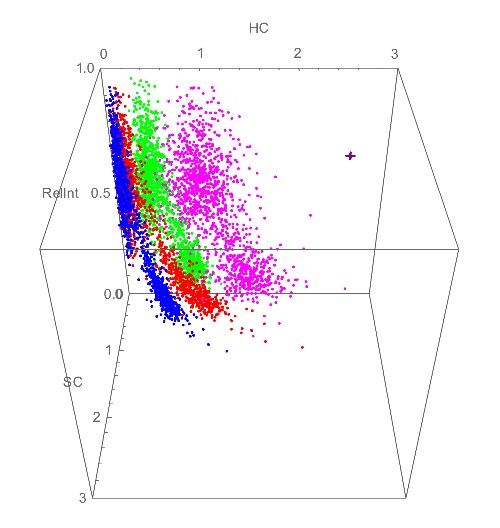}
        \subcaption{}
    \end{minipage}
    \hfill
    \begin{minipage}[t]{.2\textwidth}
        \centering
        \includegraphics[scale=0.18,angle=0]{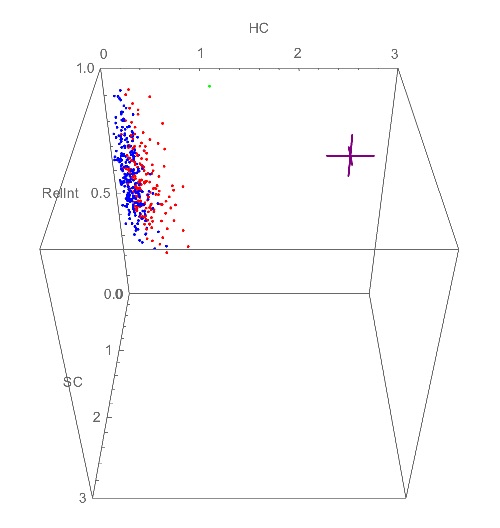}
        \subcaption{}
    \end{minipage}  
    \hfill
    \begin{minipage}[t]{.2\textwidth}
        \centering
       \includegraphics[scale=0.18,angle=0]{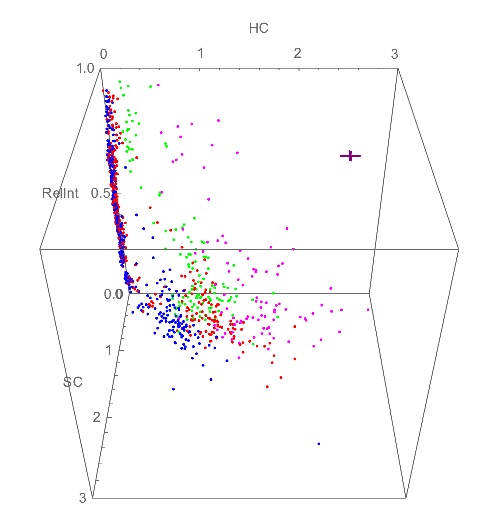}
        \subcaption{}
    \end{minipage}
    \hfill
    \begin{minipage}[t]{.2\textwidth}
        \centering
        \includegraphics[scale=0.18,angle=0]{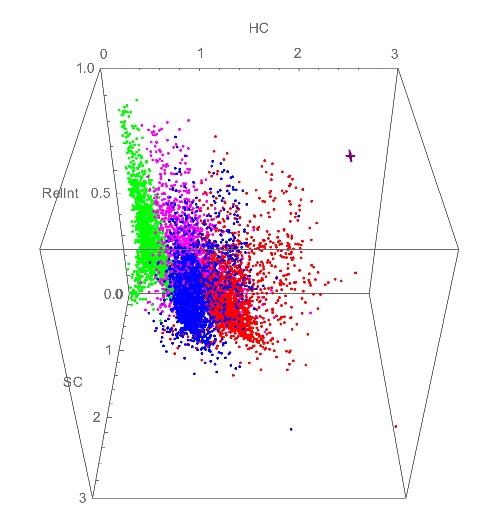}
        \subcaption{}
    \end{minipage}
    \caption{Color-Color-Intensity plots using MAXI/GSC data extracted in four of the X-ray colors listed in Table 2: CR1 (red), CR2 (blue), CR3 (magenta) and CR4 (green). The average error bar is plotted as a purple cross on each plot. {\bf (a):} Cyg X--1, a HMBH. While the location of the points in each color are slightly offset from each other, the soft and hard state are separated in each color. {\bf (b):} LMC X--1, a HMBH. Only soft state is detected by MAXI. {\bf (c):} GX339--4, a LMBH. Data includes the soft state, hard state and intermediate state. The soft state is not traced by CR3 and CR4. {\bf (d):} GRS 1915+105, a LMBH showing the intermediate spectral state in all four colors.}
    \label{compBHs}
\end{figure}

\begin{figure}[htb]
    \begin{minipage}[t]{.2\textwidth}
        \centering
       \includegraphics[scale=0.18,angle=0]{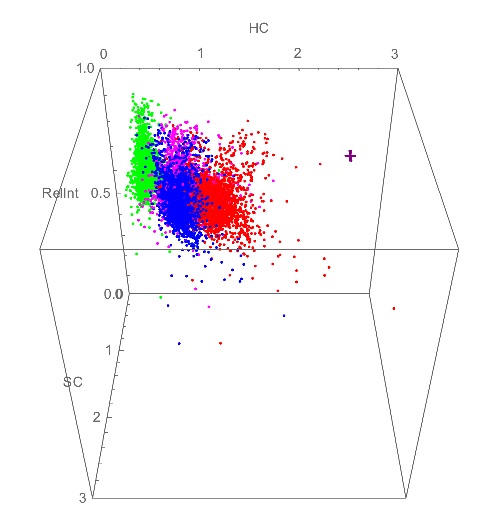}
        \subcaption{}
    \end{minipage}
    \hfill
    \begin{minipage}[t]{.2\textwidth}
        \centering
        \includegraphics[scale=0.18,angle=0]{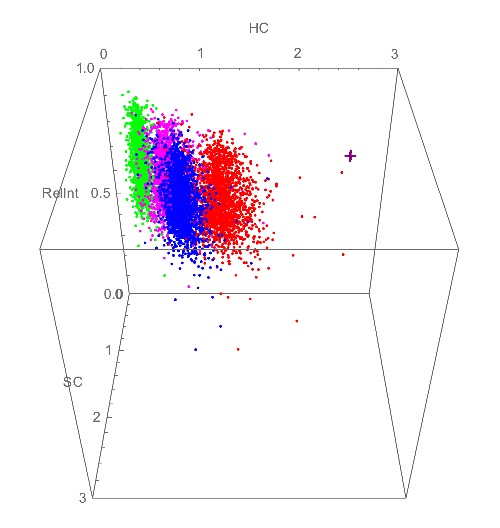}
        \subcaption{}
    \end{minipage}  
    \hfill
    \begin{minipage}[t]{.2\textwidth}
        \centering
       \includegraphics[scale=0.18,angle=0]{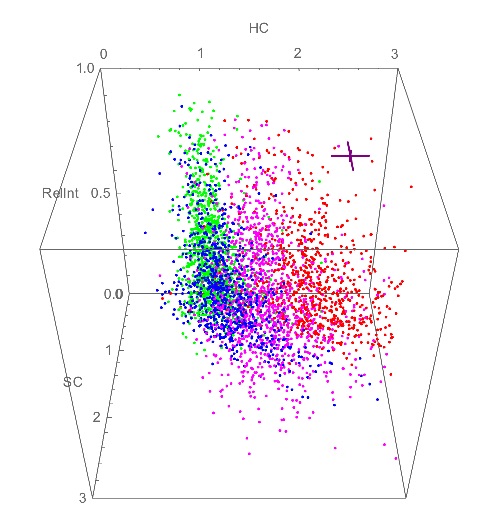}
        \subcaption{}
    \end{minipage}
    \hfill
    \begin{minipage}[t]{.2\textwidth}
        \centering
        \includegraphics[scale=0.18,angle=0]{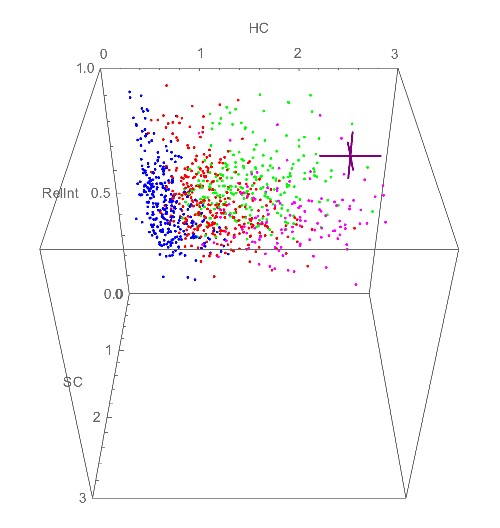}
        \subcaption{}
    \end{minipage}
    \caption{As in Figure 2 for:
{\bf (a):} GX17+2, a LMXB non-pulsing Z source
{\bf (b):} GX 9+1, a LMXB non-pulsing Atoll source; 
{\bf (c):} Cen X--3, a HMXB pulsar; 
{\bf (d):} 4U1626--67, a LMXB pulsar. While slightly offset from each other, the loci of the points of each source is traced by all four colors.}
 \label{compPNS}
\end{figure}

\begin{figure}[htb]
    \begin{minipage}[t]{.2\textwidth}
        \centering
       \includegraphics[scale=0.18,angle=0]{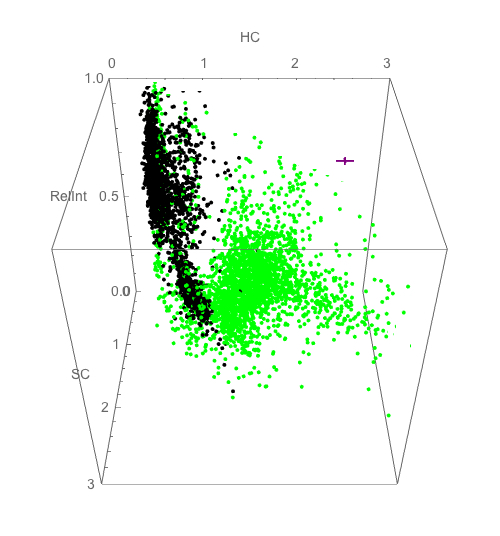}
        \subcaption{}
    \end{minipage}
    \hfill
    \begin{minipage}[t]{.2\textwidth}
        \centering
        \includegraphics[scale=0.18,angle=0]{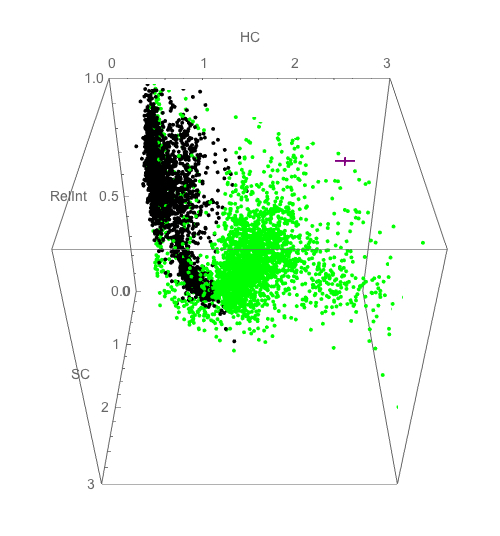}
        \subcaption{}
    \end{minipage}  
    \hfill
    \begin{minipage}[t]{.2\textwidth}
        \centering
       \includegraphics[scale=0.18,angle=0]{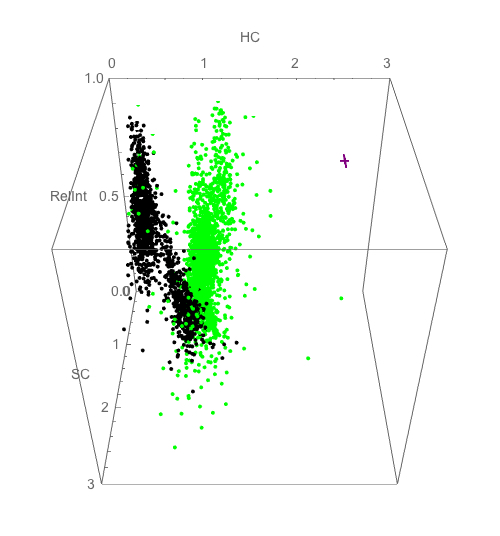}
        \subcaption{}
    \end{minipage}
    \hfill
    \begin{minipage}[t]{.2\textwidth}
        \centering
        \includegraphics[scale=0.18,angle=0]{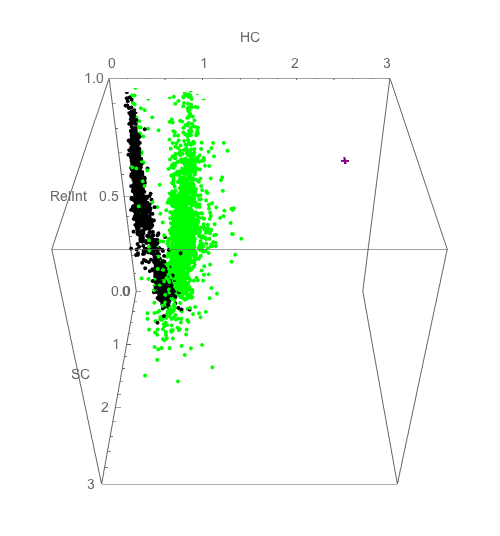}
        \subcaption{}
    \end{minipage}
    \caption{Color-Color-Intensity plots using MAXI/GSC data extracted in four of the X-ray colors listed in Table 2. The HMBHs are plotted in black and the LMBHs in green. The average error bar is plotted as purple cross on each plot. {\bf (a):} Using CR1; {\bf (b):} Using CR2; {\bf (c):} Using CR3; {\bf (d):} Using CR4. LMBHs are separated from HMBHs by all four colors. CR1 and CR2 capture the soft states whereas CR3 and CR4 do not.}
    \label{compare_BHs}
\end{figure}

\begin{figure*}
\centering
 \includegraphics[scale=0.3,angle=0]{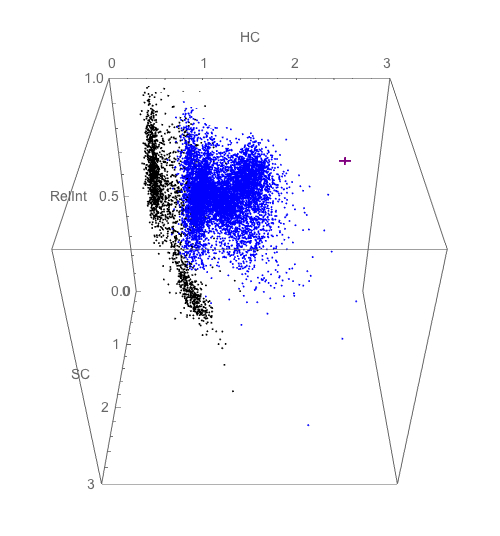}
 \includegraphics[scale=0.3,angle=0]{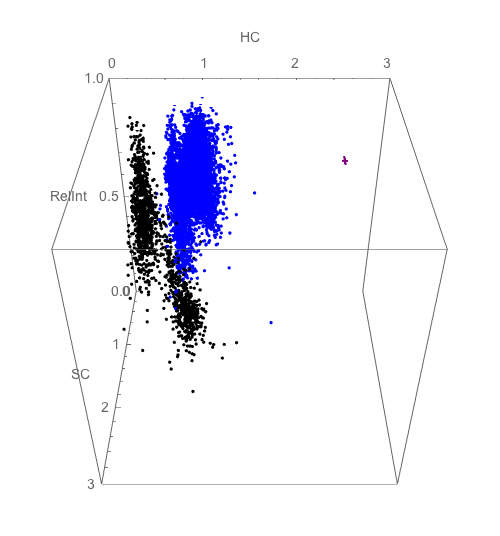}
\caption{Color-Color-Intensity plots using MAXI/GSC data extracted in CR1 (left panel)
and CR3 (right panel) for HMBHs in black and Z sources in blue. HMBHs and Z sources are separated in both colors.}
\label{compare_HMBHZ}
\end{figure*}

\begin{figure*}
 \includegraphics[scale=0.3,angle=0]{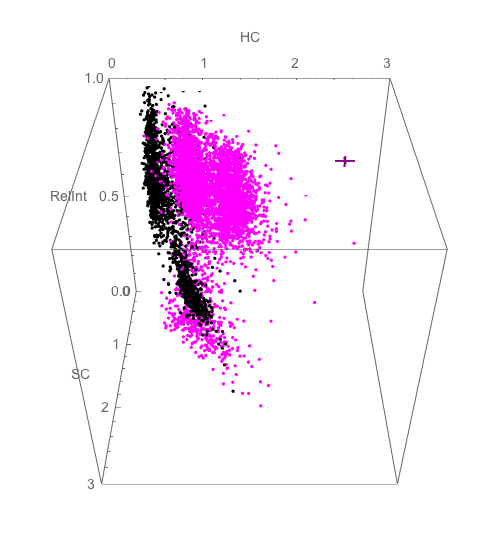}
 \includegraphics[scale=0.3,angle=0]{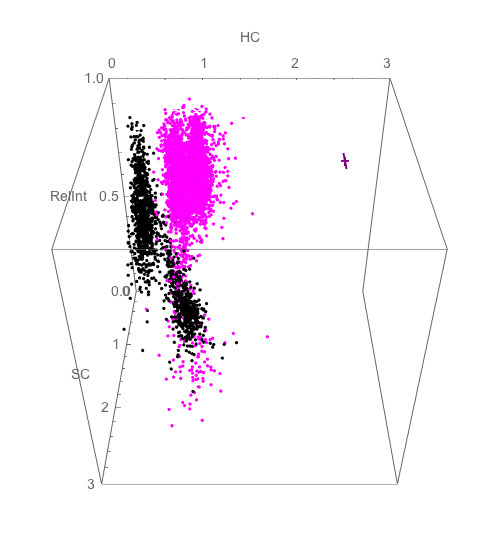}
\caption{As in Figure \ref{compare_HMBHZ} for HMBHs in black and Atoll sources in magenta. HMBHs and Atoll sources are separated in both colors.}
\label{compare_HMBHAtoll}
\end{figure*}

\begin{figure*}
 \includegraphics[scale=0.3,angle=0]{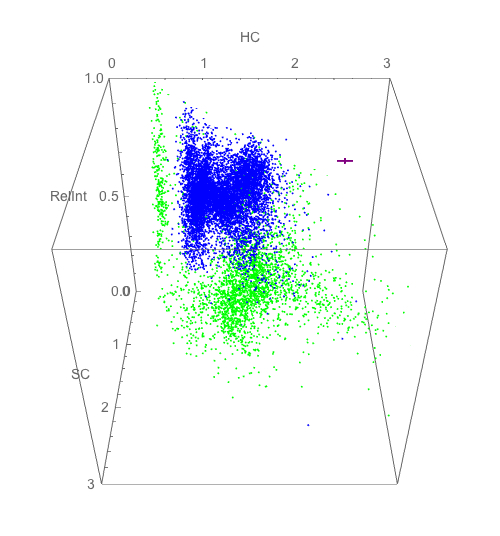}
 \includegraphics[scale=0.3,angle=0]{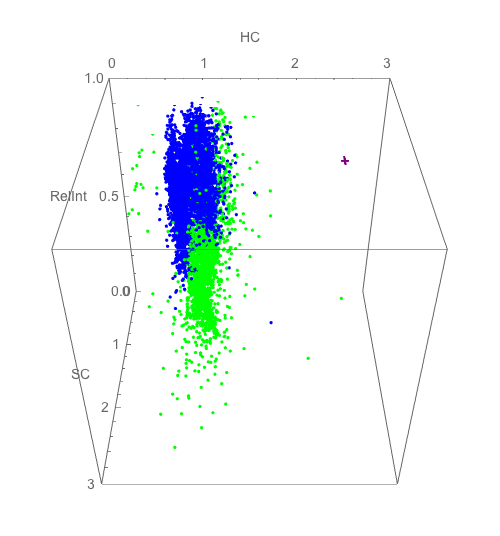}
\caption{As in Figure \ref{compare_HMBHZ} for LMBHs in green, and Z sources in blue. 
LMBHs and Atoll sources are separated in both colors.}
\label{compare_LMBHZ}
\end{figure*}

\begin{figure*}
 \includegraphics[scale=0.3,angle=0]{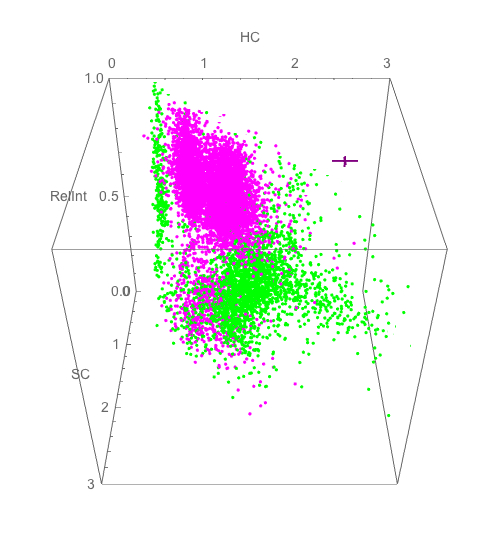}
 \includegraphics[scale=0.3,angle=0]{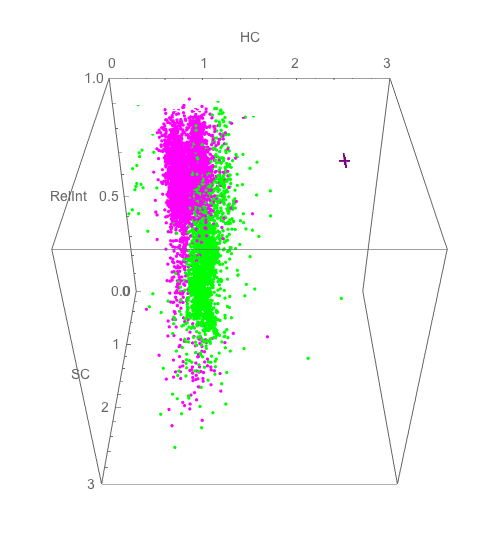}
\caption{As in Figure \ref{compare_HMBHZ} for LMBHs in green, and Atoll sources in magenta. LMBHs and Atoll sources are separated in both colors.}
\label{compare_LMBHAtoll}
\end{figure*}

\begin{figure*}
 \includegraphics[scale=0.3,angle=0]{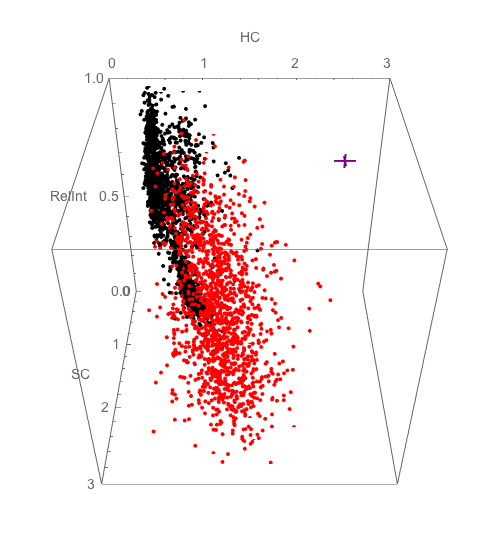}
 \includegraphics[scale=0.3,angle=0]{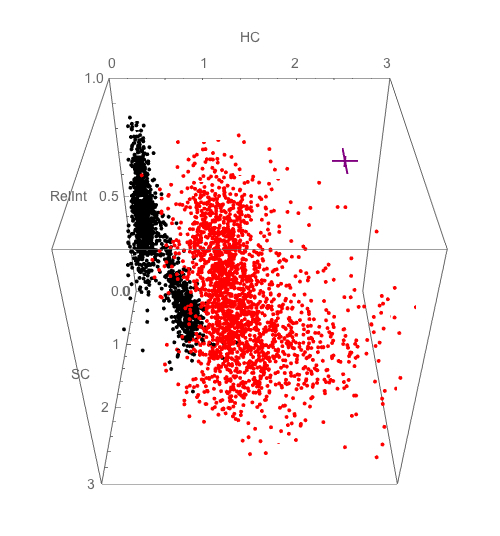}
\caption{As in Figure \ref{compare_HMBHZ} for HMBHs in black, and HMXB pulsars in red. 
HMBHs and HMXB pulsars are separated in both colors.}
\label{compare_HMBHHMpuls}
\end{figure*}

\begin{figure*}
 \includegraphics[scale=0.3,angle=0]{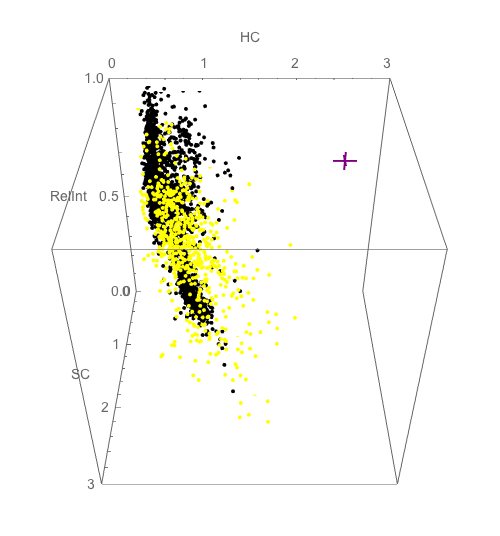}
 \includegraphics[scale=0.3,angle=0]{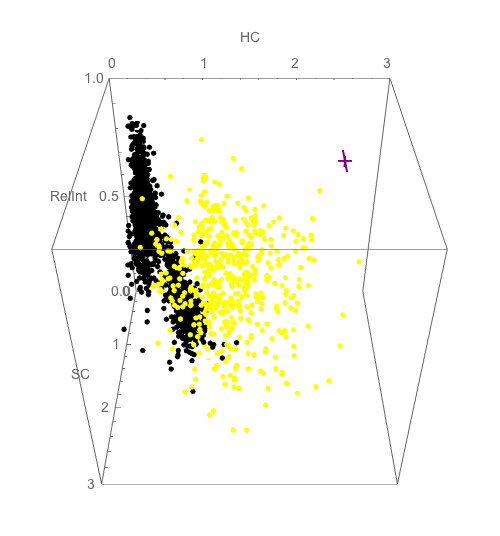}
\caption{As in Figure \ref{compare_HMBHZ} for HMBHs in black, and LMXB pulsars in yellow. HMBHs and LMXB pulsars are separated in both colors.}
\label{compare_HMBHLMpuls}
\end{figure*}

\begin{figure*}
 \includegraphics[scale=0.3,angle=0]{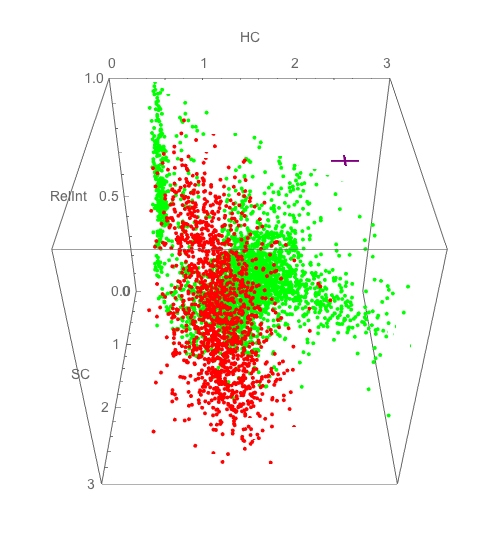}
 \includegraphics[scale=0.3,angle=0]{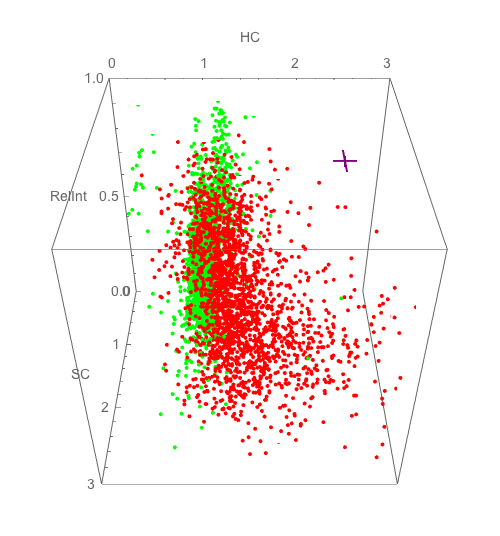}
\caption{As in Figure \ref{compare_HMBHZ} for LMBHs in green, and HMXB pulsars in red.
 LMBHs and HMXB pulsars are separated in both colors.}
\label{compare_LMBHHMpuls}
\end{figure*}

\begin{figure*}
 \includegraphics[scale=0.3,angle=0]{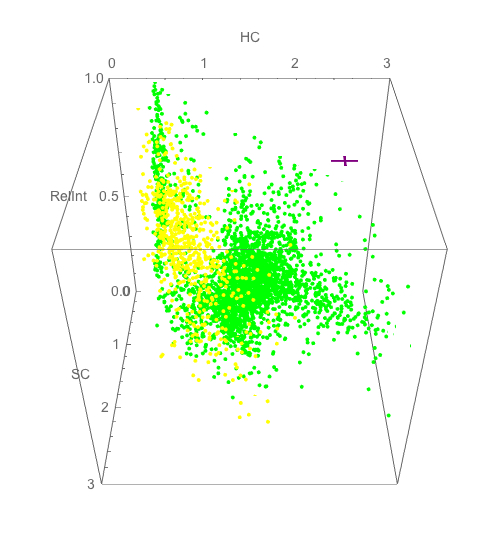}
 \includegraphics[scale=0.3,angle=0]{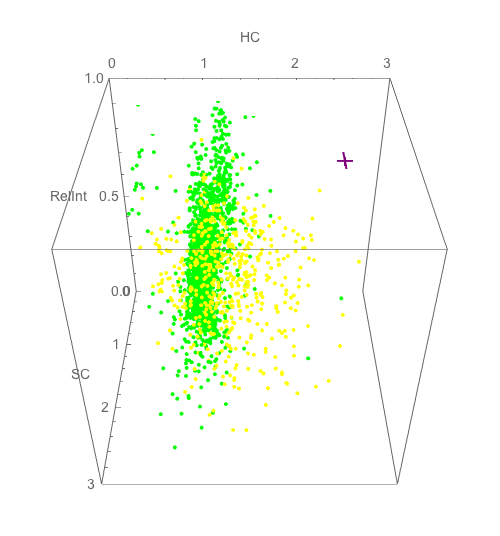}
\caption{As in Figure \ref{compare_HMBHZ} for LMBHs in green, and for LMXB pulsars in yellow. LMBHs and LMXB pulsars are separated in both colors. }
\label{compare_LMBHLMpuls}
\end{figure*}

\begin{figure*}
 \includegraphics[scale=0.3,angle=0]{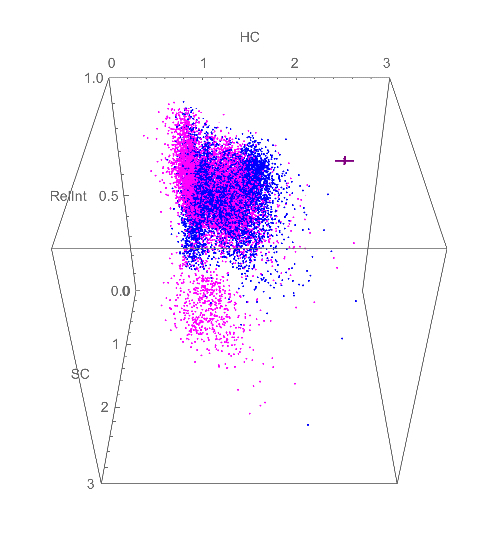}
 \includegraphics[scale=0.3,angle=0]{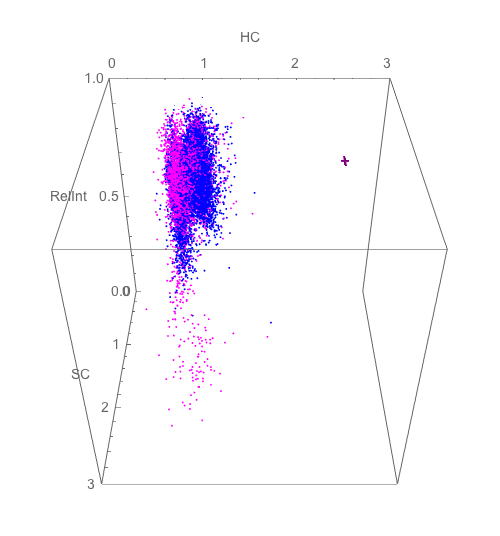}
\caption{As in Figure \ref{compare_HMBHZ} for Z sources in blue, and for Atoll sources in magenta. Atoll and Z sources cannot be distinguished.}
\label{compare_ZAtoll}
\end{figure*}

\begin{figure*}
\includegraphics[scale=0.3,angle=0]{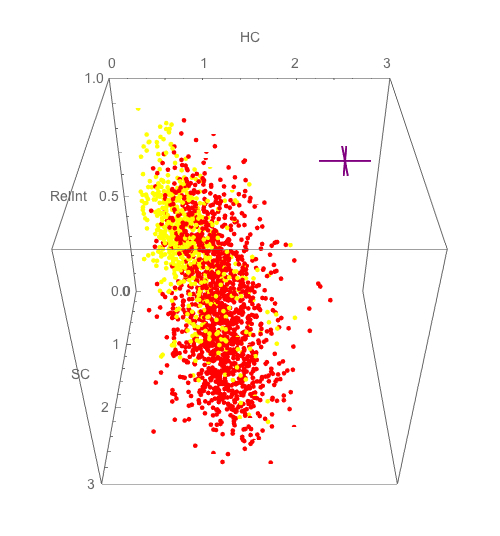}
\includegraphics[scale=0.3,angle=0]{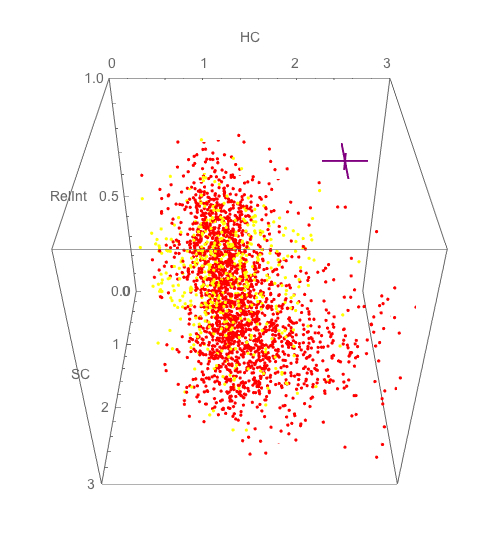}
\caption{As in Figure \ref{compare_HMBHZ} for HMXB pulsars in red, and LMXB pulsars in yellow. It shows some overlap between HMXB and LMXB pulsars.}
\label{compare_HMLMpuls}
\end{figure*}

\begin{figure*}
 \includegraphics[scale=0.3,angle=0]{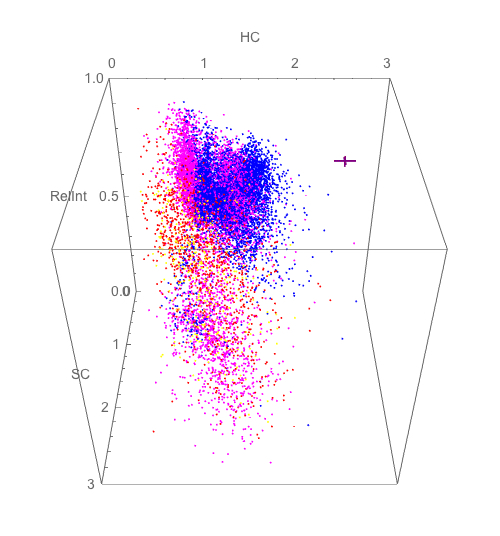}
 \includegraphics[scale=0.3,angle=0]{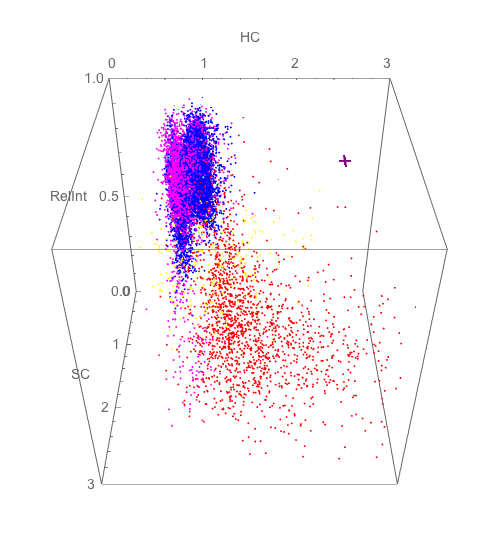}
\caption{As in Figure \ref{compare_HMBHZ} for Z sources in blue, Atoll sources in magenta,  HMXB pulsars in red, and LMXB pulsars in yellow. }
\label{compare_ZHMPulsar}
\end{figure*}

\begin{figure}[htb]
    \begin{minipage}[t]{.2\textwidth}
        \centering
       \includegraphics[scale=0.18,angle=0]{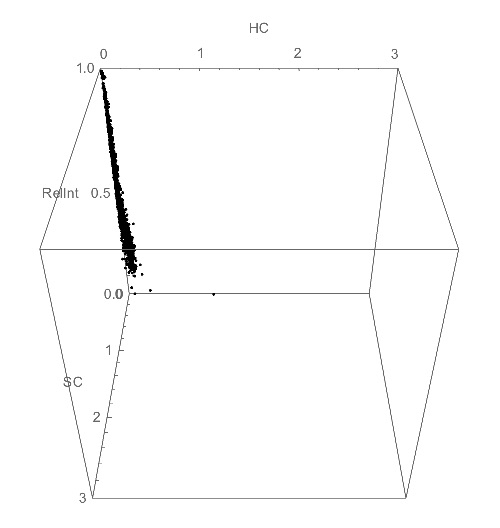}
        \subcaption{}
    \end{minipage}
    \hfill
    \begin{minipage}[t]{.2\textwidth}
        \centering
        \includegraphics[scale=0.18,angle=0]{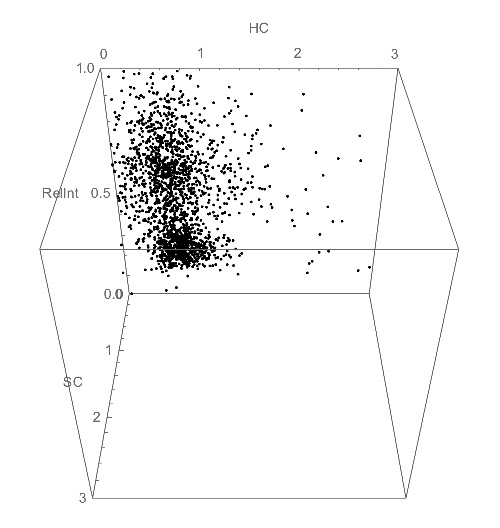}
        \subcaption{}
    \end{minipage}  
    \hfill
    \begin{minipage}[t]{.2\textwidth}
        \centering
       \includegraphics[scale=0.18,angle=0]{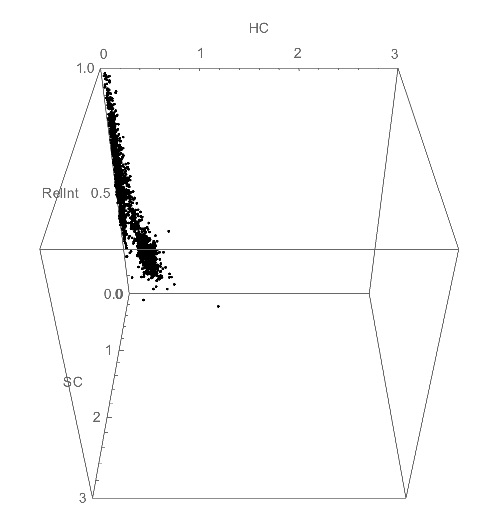}
        \subcaption{}
    \end{minipage}
    \hfill
    \begin{minipage}[t]{.2\textwidth}
        \centering
        \includegraphics[scale=0.18,angle=0]{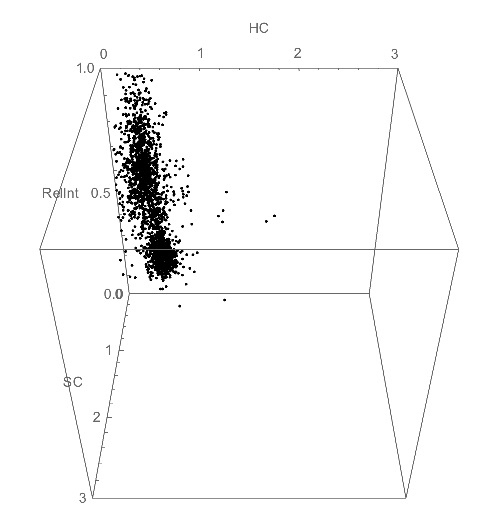}
        \subcaption{}
    \end{minipage}
    \caption{Color-Color-Intensity plots using MAXI/GSC data of the HMBH Cyg X--1 extracted in each of the X-ray colors {\bf (a):} CR5; {\bf (b):} CR6; {\bf (c):} CR7; and {\bf (d):} CR8 as listed in Table 2. CR5 and CR7 are better at separating the hard and soft states. }
 \label{compare_cygx1}
\end{figure}

\begin{figure}[htb]
    \begin{minipage}[t]{.2\textwidth}
        \centering
       \includegraphics[scale=0.18,angle=0]{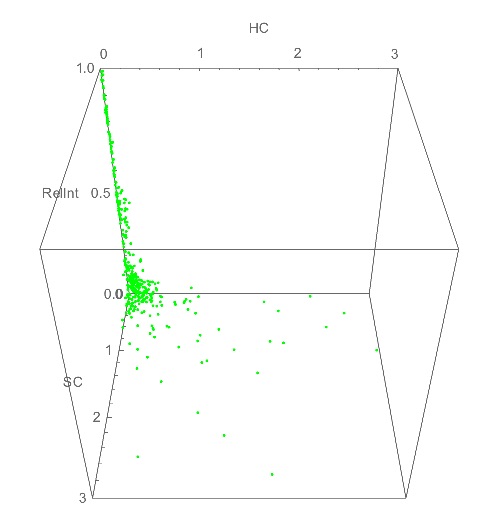}
        \subcaption{}
    \end{minipage}
    \hfill
    \begin{minipage}[t]{.2\textwidth}
        \centering
        \includegraphics[scale=0.18,angle=0]{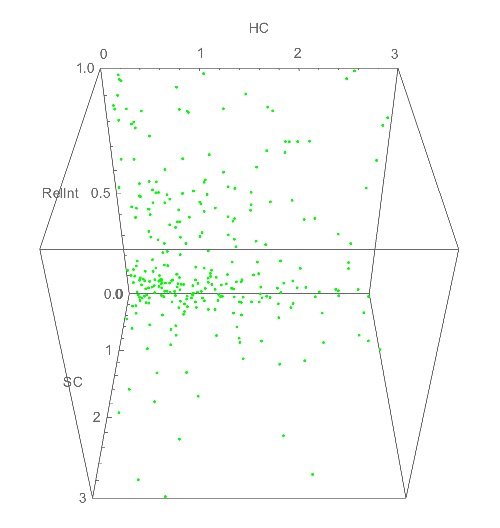}
        \subcaption{}
    \end{minipage}  
    \hfill
    \begin{minipage}[t]{.2\textwidth}
        \centering
       \includegraphics[scale=0.18,angle=0]{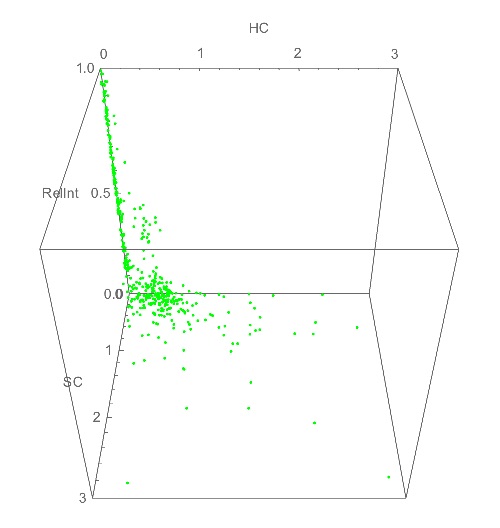}
        \subcaption{}
    \end{minipage}
    \hfill
    \begin{minipage}[t]{.2\textwidth}
        \centering
        \includegraphics[scale=0.18,angle=0]{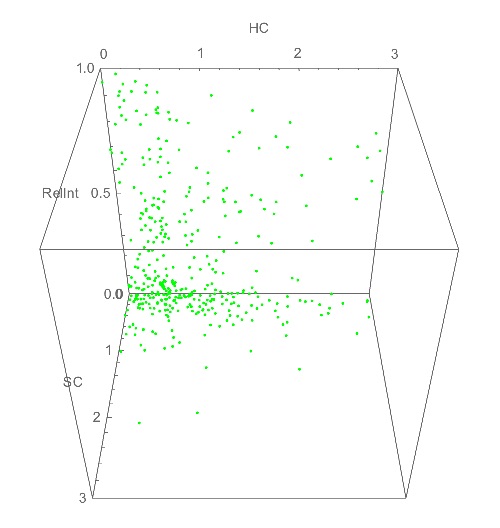}
        \subcaption{}
    \end{minipage}
    \caption{Color-Color-Intensity plots using MAXI/GSC data of the LMBH GX 339--4 extracted in each of the X-ray colors {\bf (a):} CR5; {\bf (b):} CR6; {\bf (c):} CR7; and {\bf (d):} CR8 as listed in Table 2. Only CR7 suggests the possibility of three states in the system (soft, hard, and intermediate).}
 \label{compare_gx339}
\end{figure}

\begin{figure}[htb]
    \begin{minipage}[t]{.2\textwidth}
        \centering
       \includegraphics[scale=0.18,angle=0]{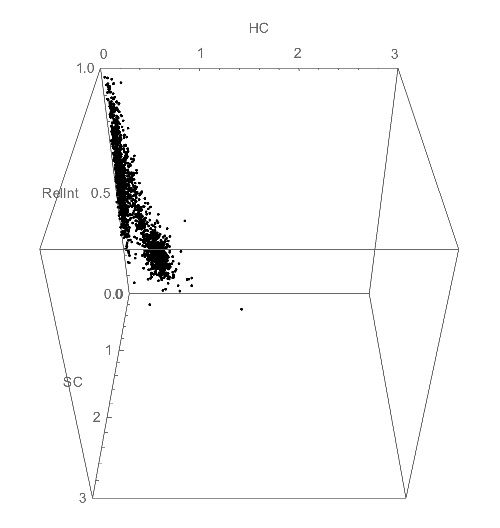}
        \subcaption{}
    \end{minipage}
    \hfill
    \begin{minipage}[t]{.2\textwidth}
        \centering
        \includegraphics[scale=0.18,angle=0]{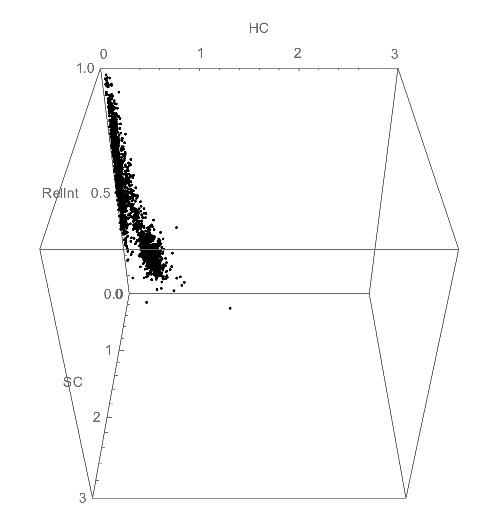}
        \subcaption{}
    \end{minipage}  
    \hfill
    \begin{minipage}[t]{.2\textwidth}
        \centering
       \includegraphics[scale=0.18,angle=0]{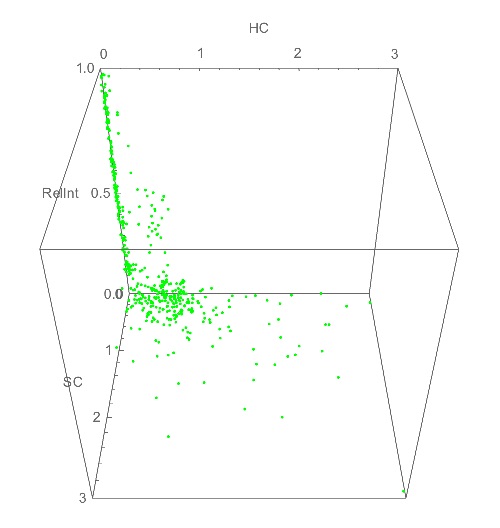}
        \subcaption{}
    \end{minipage}
    \hfill
    \begin{minipage}[t]{.2\textwidth}
        \centering
        \includegraphics[scale=0.18,angle=0]{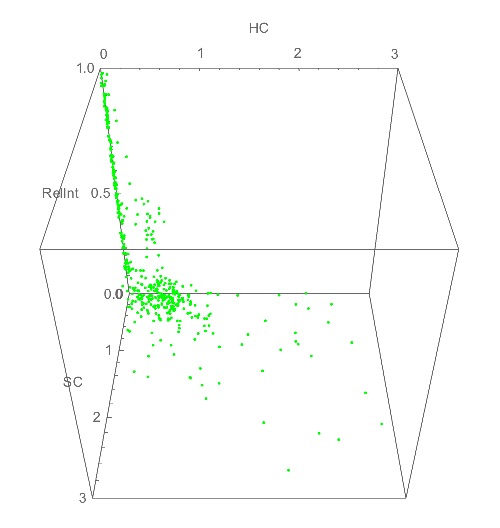}
        \subcaption{}
    \end{minipage}
    \caption{Color-Color-Intensity plots using MAXI/GSC data extracted in each of
the X-ray colors (left to right panels): {\bf (a):} CR9 for Cyg X-–1; {\bf (b):} CR10 for Cyg X--1; {\bf (c):} CR9 for GX 339--4; and {\bf (d):} CR10 for GX 339--4. Both colors separate hard and soft states and where present, the intermediate state.}
 \label{compare_3vs4}
\end{figure}

\begin{figure*}
\centering
\includegraphics[scale=0.3,angle=0]{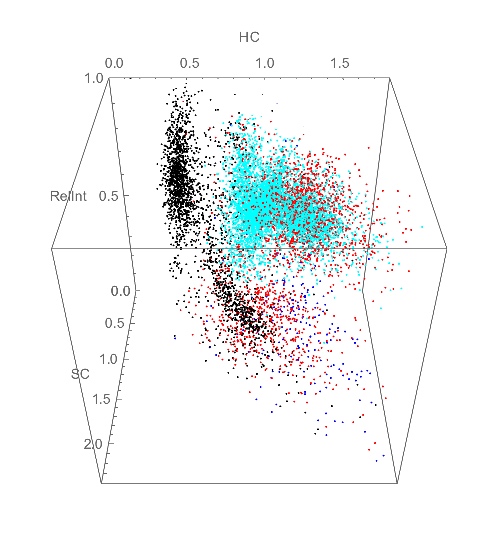}
\includegraphics[scale=0.3,angle=0]{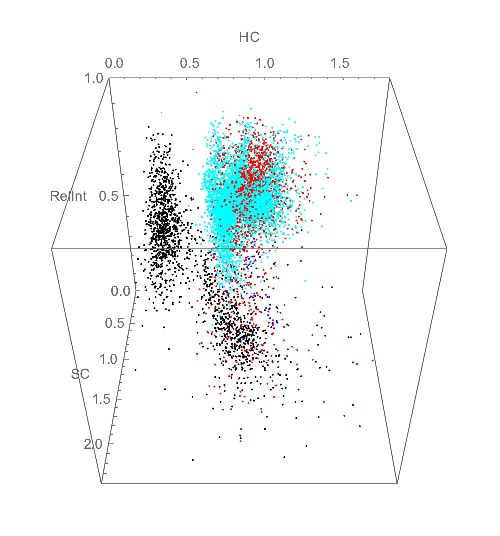}
\caption{Color-Color-Intensity plots using MAXI/GSC data extracted for all the
XRBs plotted in (\citep{done2003}; their Fig. 8) shown for CR1 (left panel) and CR3 (right panel). Our plots extend to the same HC and SC limits that they used and we use a similar color scheme BHs are in black, Z sources in red, Atoll sources in cyan, and Cir X–-1 in
blue. The region \cite{done2003} identify as inaccessible to NS is clearly isolated. In CCI space (when seen in 3 dimensions) colors CR1 and CR2 completely separate all BH
systems from all systems containing NS. Cir X--1 shows no overlap with BHs.}
\label{NSvsBH}
\end{figure*}

\subsection{Color-Color-Intensity Diagrams}

In CCI diagrams, one coordinate referred to as the Soft Color (SC) is a ratio involving the softest available energy band, the second coordinate referred to as the Hard Color (HC) 
is a ratio involving the highest available energy band, and the third coordinate is a sum of the counts in all energy bands and represents the source intensity (INT).
In this paper, the third coordinate is normalized by dividing the total counts by the average of the top 1\% of the data for any given source, hence we plot the relative intensity (RelInt).  
\par
For X-ray colors  which use four energy bands, SC is the ratio of the two lower energy bands and HC is the ratio of the two higher energy bands. For X-ray colors which use three energy bands, SC is the ratio of the middle energy band to the lowest energy band and HC is the ratio of the high energy band to the lowest energy band (see Table 2)
\par
All the plots used in this paper are available at a website\footnote{\url{http://hea-www.harvard.edu/\~saku/maxilinks.html}} as moving gifs so that readers are able to see the 3 dimensional structure around the full 360$^{\circ}$. It is essential to see the 3-D images since 2-D projections can show overlap when separation occurs in the third dimension.

\section{Effects of different X-ray colors on separation of XRB classes}

We first use colors that have been previously defined in the literature to construct CCI plots of sources listed in Table 1. The colors are defined over a range of energies that are sub-divided into three or four energy-bands. As itemized in Table 2, CR1 and CR2 are defined for energies ranging up to 8 keV and 12 keV and separated into three energy-bands, CR3 and CR4 are defined for energy-bands extending up to 18 keV and separated into four energy-
bands. In these plots, CR1 is depicted in red, CR2 in blue, CR3 in magenta, and CR4 in green. An average error bar is plotted as a purple cross on each plot.
\par
For Cyg X--1, a HMBH (Figure \ref{compBHs}a), the high soft state (within each color: points lying at low HC and SC and at high RelInt values), and the low hard state (within each color: points lying at higher HC and SC and lower RelInt values) are separated in all four colors. The points are more localized for CR1 (red) and CR2 (blue) and show a larger spread for CR3 and CR4 (magenta and green). LMC X--1 (Figure \ref{compBHs}b), is also a HMBH, but since it is extragalactic, the low hard state is not accessible to MAXI/GSC and only the high soft state is observed. We note that LMC X--1 is not visible for CR3 and CR4 (magenta and green) because it has little or no counts in the hard energy bands used to define HC in those colors. GX 339--4, A LMBH (Figure \ref{compBHs}c), shows the high soft state (points close to zero for SC and HC and higher RelInt), the low hard state (points with high SC and HC and low RelInt), as well as points intermediate between the two in CR1 and CR2 (red and blue); only the intermediate and low hard states are detected by CR3 and CR4 (magenta and green). GRS 1915+105 (Figure \ref{compBHs}d), is also a LMBH, but it has been in outburst since 1992 and its spectra show only intermediate states \citep{belloni1997} which are detected in all four X-ray colors.
\par
Figure \ref{compPNS} show the CCI plots for one each of the sample of NS XRBs listed in Table 1. The CCI plots for NS XRBs not shown are similar to the ones displayed for each class. The position of a source is shifted for different X-ray colors and has different degrees of scatter.  The HMXB pulsars are wind-fed systems and show considerably more intensity variations than the LMXB pulsars.
\par
Next we consider different classes of XRBs, pairwise, in each of CR1, CR2, CR3, and CR4. 
 For these Figures (\ref{compare_BHs} -- \ref{compare_ZHMPulsar}), HMBHs are depicted in black; LMBHs in green; Z sources in blue; Atoll sources in magenta; HMXB pulsars in red; and LMXB pulsars in yellow.
\par
Figure \ref{compare_BHs} shows the CCI plots for the sample of HMBHs (black) and LMBHs (green) for CR1--CR4.  For CR1 (Figure \ref{compare_BHs}a) and CR2 (Figure \ref{compare_BHs}b), there is some overlap between the high soft states of HMBH and LMBH. CR3 (Figure \ref{compare_BHs}c) and CR4 (Figure \ref{compare_BHs}d) distinguish between HMBH and LMBH but do not detect the soft states.  We investigate the effects of X-ray colors on separation of states in detail in Section 3.1.
 \par
 Since CR1 is very similar to CR2, and CR3 is very similar to CR4, we only show the plots corresponding to CR1 and CR3 in the rest of the paper. CCI plots for CR2 and and CR4 are included on the website.\footnote{\url{http://hea-www.harvard.edu/\~saku/maxilinks.html}} We compare non-pulsing Z sources (blue) with that of HMBHs (black; Figure \ref{compare_HMBHZ}). The non-pulsing Z sources have higher HC than the soft states of HMBHs and they have higher RelInt than the hard states of HMBHs in all four X-ray colors. Therefore, Z sources are clearly separated from the HMBHs in all four colors. 
\par
Figure \ref{compare_HMBHAtoll} compares the sample of non-pulsing NS Atoll sources (magenta) with that of HMBHs (black). The Atoll sources also have higher HC than the soft states of HMBHs, and higher RelInt than the hard states of HMXBs. Therefore, the Atoll sources are separated from HMBHs in all four colors. 
\par
Figures \ref{compare_LMBHZ} and \ref{compare_LMBHAtoll} show that non-pulsing NS systems (both Z and Atoll sources) are easily separated from LMBHs in all four X-ray colors CR1--CR4, since both Z and Atoll sources have higher HC than soft states of LMBHs and higher RelInt than hard states of LMBHs, and occupy a different plane than the intermediate state of LMBHs.
\par
We next compare the sample of NS pulsars with that of BH binaries. Figures \ref{compare_HMBHHMpuls} and \ref{compare_HMBHLMpuls} compare HMBHs (black) with HMXB pulsars (red) and LMXB pulsars (yellow) respectively. Figures \ref{compare_LMBHHMpuls} and \ref{compare_LMBHLMpuls} compare LMBHs (green) with HMXB pulsars (red) and LMXB pulsars (yellow) respectively.
The pulsing neutron star systems (both HMXB and LMXB) have higher HCs than that of the BHs (both HMBHs and LMBHs); so they are also distinguished in all four colors.
\par
We compare the sample of Z and Atoll sources in Figure \ref{compare_ZAtoll}. We find these two classes of non-pulsing neutron star X-ray binaries occupy similar regions showing considerable overlap in all tested colors (CR1, CR2, CR3, CR4); however, Atolls do show a low state that does not appear in Z sources. Figure \ref{compare_HMLMpuls} shows that there is considerable overlap between HMXB and LMXB pulsars. We discuss these results in Section 3.  Since Z and Atoll sources cannot be distinguished from each other, and HMXB and LMXB pulsars cannot be distinguished from each other, we plot them together in order to compare the effect of colors in separating pulsing vs non-pulsing systems (Fig. \ref{compare_ZHMPulsar}).  All colors distinguish non-pulsing NS systems from pulsing NS systems.

\section{Effects of different X-ray colors on separation of spectral states of BH binaries}
In Section 3, we found that CR3 and CR4 did not detect the soft spectral state of BH XRBs. CR1 and CR2 are defined for three energy-bands and extend to 8 and 12 keV, whereas CR3 and CR4 are colors defined for four energy-bands and extend up to 18 keV.  We define colors CR5--CR8 (listed in Table 2), to cover the maximum energy range (2-18 keV) in different selections of three- and four-energy bands to further investigate the effects of different X-ray colors on the separation of spectral states of BH XRBs. CR5, CR7 are defined for three energy-bands whereas CR6 and CR8 are defined for four energy-bands.
\par
Figures \ref{compare_cygx1} and \ref{compare_gx339} show Cyg X--1 (in black) and GX 339--4 (in green) for colors CR5, CR6, CR7, and CR8 (left to right panels). There is significantly higher scatter in these plots for the colors defined with four energy-bands (CR6 and CR8) than the colors defined with three energy-bands (CR5, CR7). Also the CCI diagrams that use three-color bands (Figure \ref{compare_cygx1} a,c and Figure  \ref{compare_gx339} a,c) separate the soft and hard states of Cyg X--1 and GX 339--4 (visible only in 3D for CR5) whereas the diagrams using four-color bands (Figure \ref{compare_cygx1} b,d and Figure \ref{compare_gx339} b,d) do not.  
\par
In order to test if it is the ratio of energy bands and not the energy range alone that causes this behavior, we defined CR9 and CR10 for the exact same energy ranges as 
CR3 and CR4 but using three energy bands instead of four.  Figure \ref{compare_3vs4} a,b and Figure \ref{compare_3vs4} c,d show that both CR9 and CR10 are able to separate states for both GX 339--4 and Cyg X--1.  The exact same energy ranges when divided into four energy bands (as we observed earlier in Figures 4c and 4d) do not separate the spectral states. In the definitions using three X-ray bands the hard and soft colors are both normalized to the lowest energy-band; in definitions using four energy bands the hard color is defined only for the highest energy bands; since BHs are mostly soft sources the high energy bands contribute very little information.

\section{Using CCI diagrams for classifying unknown XRB type}

\begin{figure*}
 \includegraphics[scale=0.3,angle=0]{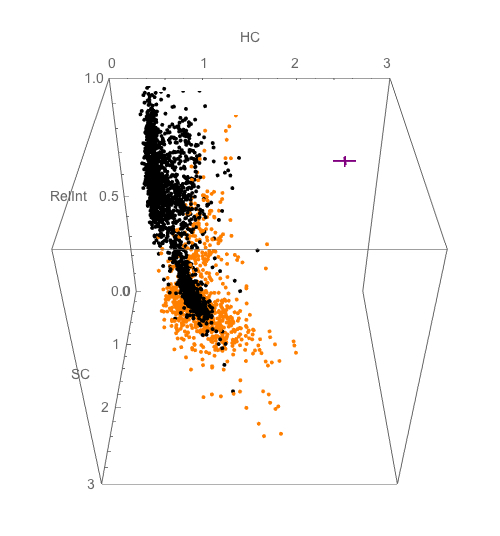}
 \includegraphics[scale=0.3,angle=0]{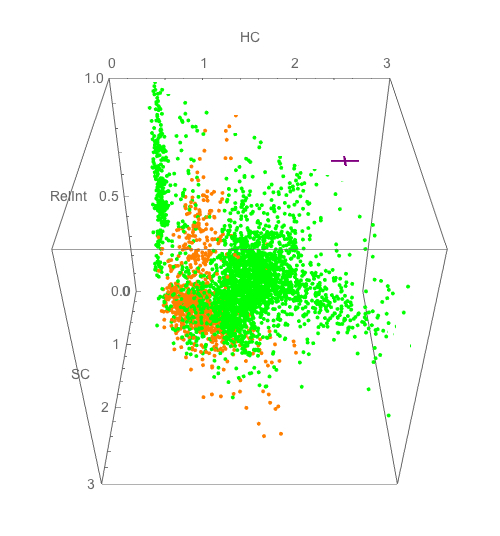}
\caption{Color-Color-Intensity plots using MAXI/GSC data with HMBHs in black, LMBHs in green and GRS 1739--278 in orange. The data-points for GRS 1739--278 are located closer to LMBHs than to HMBHs, identifying it as a LMBH.}
\label{compare_grs1739}
\end{figure*}

\begin{figure*}
 \includegraphics[scale=0.3,angle=0]{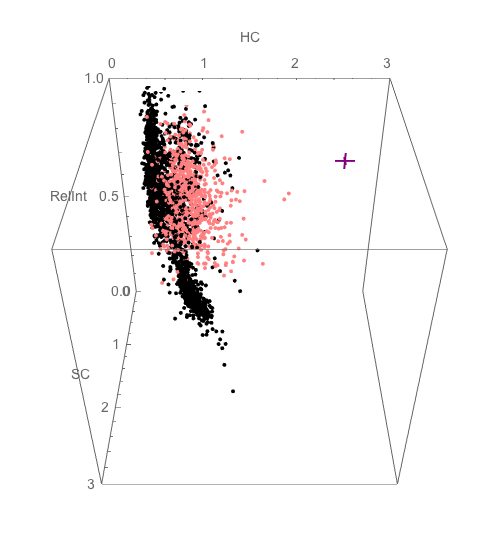}
 \includegraphics[scale=0.3,angle=0]{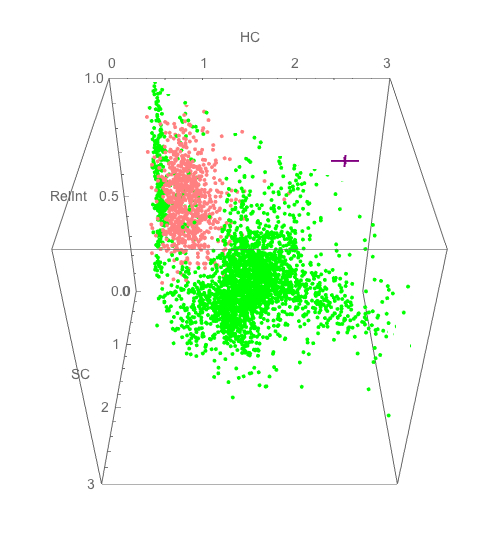}
\caption{Color-Color-Intensity plots using MAXI/GSC data with HMBHs in black, LMBHs in green and 4U 1957+115 in pink. The data-points for 4U 1957+115 show an overlap with both HMBHs and LMBHs in their high soft states.}
\label{compare_4u1957}
\end{figure*}

We use CCI diagrams to classify the XRBs GRS 1739--278 and 4U 1957+115 on the basis of their locations compared to known sources. GRS 1739--278 is a transient  \citep{furst2016} and 4U 1957+115 is always active in X-rays \citep{nowak2008}. We compared the CCI diagrams of GRS 1739--278 and 4U 1957+115 with all classes of XRBs using CR1. We show the comparison with the HMBHs and LMBHs in Figures \ref{compare_grs1739} and \ref{compare_4u1957}.  Plots from comparison with the other XRB classes are included on the website.\footnote{\url{http://hea-www.harvard.edu/\~saku/maxilinks.html}}
\par
The data-points of GRS 1739--278 are separated from the Z sources, and both HM and LM pulsars, however they show some overlap with the low states of Atoll sources. As seen in Figure \ref{compare_grs1739}, the data-points for GRS 1739--278 are separated from the HMBHs. There is significant overlap with LMBHs, hence we classify GRS1739--278 as a LMBH. This is consistent with the spectral and timing characteristics of GRS 1739--278 as described by \cite{furst2016}.  
\par
The data-points of 4U 1957+115 are completely disjoint from the HM and LM pulsars. However, they show some overlap with Z and Atoll sources.  As seen in Figure \ref{compare_4u1957}, the data-points for 4U 1957+115 show an overlap with both HMBHs and LMBHs in their high soft states. This is consistent with \cite{nowak2008} who found 4U 1957+115 to be in a persistent soft state. We can conclude that 4U 1957+115 is indeed a black hole but there is ambiguity in its classification as either a HMBH or LMBH based on CCI diagrams.

\section{Possible physics driving systems to given locations in CCI diagrams}
DG03 studied accretion behavior in BHs and NSs as a function of Eddington luminosity. They restricted themselves to sources that show high variability in their bright states (hence cover a wide range of spectral states) and low interstellar absorption (so that X-ray colors are robust). They selected 14 XRBs that had significant amounts of RXTE/PCA data (Table 1 in DG03). They used masses and distances from the literature to convert observed fluxes into L/L$_{Edd}$. DG03 showed that the evolution of a source as a function of luminosity varied based on the type of compact object: there is an area in the CC plot (Figure 8 in \citealt{done2003}) containing only BH systems. They attribute this area to the fact that BHs do not have solid surfaces; NS systems cannot occupy this region because of the additional emission from their boundary-layer. They also connected this area with a type of X-ray spectrum that is detected only from BHs and suggest this spectrum as a diagnostic to determine the nature of new transient sources. However, there are significant areas in their CC plot where systems containing NSs and BHs overlap.
\par
We extracted 12 of the sources studied by DG03 with MAXI/GSC data in each of X-ray colors CR1, CR2, CR3, CR4 (the other two sources, XTE J1550--564 and XTE J1859+226, did not go into a bright state during MAXI observations). We display the same ranges in X-ray color that DG03 use (HC from 0.0-1.8 and SC from 0-2.4). Figure \ref{NSvsBH} shows these CCI plots in CR1 and CR3. 
We use the color-scheme from Figure 8 of DG03: black for BHs, red for Z sources, cyan for Atoll sources, blue for Cir X--1, for ease of comparison. We find that not only do CCI diagrams show the region occupied only by BHs; all systems containing BHs are separated from all systems containing NSs. Hence CCI diagrams provide a simpler and more reliable means than DG03 of determining the nature of the compact object in hitherto unclassified XRBs. The overlap between Z sources and Atoll sources can be attributed to the mass accretion rate: as \cite{fridriksson2015} point out: the same source can show Z shapes in CC diagrams when at high mass accretion rate and Atoll shapes when at low mass accretion rate.
\par
Finally, we note that CCI diagrams also clearly separate out X-ray pulsars from non-pulsing NS and BHs (Figure 5 to Figure 15). As \cite{massi2008} suggest this separation is due the strength of the magnetic field. Non pulsing NS and BHs with weak magnetic fields are totally separated from pulsars, which have strong magnetic fields. This is likely because the dominant X-ray emission from pulsars originates at the poles whereas the non-pulsing sources are dominated by softer emission from the surface of the NS.

\section{Discussion and Conclusions}

Previous work by VB13 constructed CCI plots for 48 XRBs using 13 years of data in the predefined X-ray colors available from RXTE/ASM.  VB13 found that the various classes (systems containing black holes, pulsing and non-pulsing neutron stars) occupy distinct regions in CCI diagrams.
\par
In this paper, we expand on the work of VB13 using a more sensitive instrument (MAXI/GSC) and for a range of energy bands defining ten different sets of X-ray colors (CR1 -- CR10). We also test the ability of CCI diagrams to distinguish spectral states of BH XRBs. Since the lower energy band of MAXI/GSC is 2 keV, it is less sensitive to effects of  absorption and scattering from the interstellar matter and the circumstellar matter present around these XRBs.
\par
We confirm the ability of CCI diagrams to differentiate the types of compact object contained in a XRB (Figures \ref{compare_BHs} -- \ref{compare_ZHMPulsar}). Systems containing BHs (both HMBH and LMBH) are clearly separated from the systems containing NSs (both pulsing and non-pulsing) in colors CR1 -- CR4. HMBHs are easily distinguished from LMBHs in colors CR3 and CR4, however CR1 and CR2 show some overlap in the high soft state. The non-pulsing NS systems (Z and Atoll sources) can be distinguished from
both HMXB and LMXB pulsars in all four colors. We confirm the result of VB13 that there is considerable overlap between Z and Atoll NS systems in CCI diagrams; this is consistent with  \cite{homan2010} and \cite{fridriksson2015} conclusion that the specific patterns of a given source are dependent on mass accretion rate. We also found considerable overlap between HMXB and LMXB pulsars which suggests that the effect of the high magnetic fields in these systems outweighs the effect of the masses of their companions on their
position in CCI diagrams. However, there is a significant difference between the range of fluxes displayed by HMXB pulsars and LMXB pulsars so they should be distinguishable using the machine learning techniques of \cite{gopalan2015} and and de Beurs et al (ApJ, submitted).
\par
Within a specified energy range, we find that certain colors are better than others at detecting the soft states of BH binaries. In particular we found that all the colors tested that were formed using three energy bands (CR1, CR2, CR5, CR7, CR9, and CR10) are able to separate spectral states whereas none of the four energy band colors tested (CR3, CR4, CR6, CR8) did (Figures \ref{compare_BHs}, \ref{compare_cygx1}, \ref{compare_gx339} and \ref{compare_3vs4}; see \url{http://hea-www.harvard.edu/\~saku/maxilinks.html} for 3-D figures). This is likely because in the definitions that use four energy bands the HC is defined for the two highest energy bands and these bands have very few counts, if any, during soft states. 
\par
Finally the separation between BHs and NSs can be attributed to the presence of a boundary layer in NSs and the absence of surfaces on BHs as suggested by DG13. DG13 defined this area with detailed spectral analysis of different regions in CI diagrams; they found that certain regions were associated with a specific type of X-ray spectrum found only in BHs. They suggest that fits to unknown sources that required this type of X-ray spectrum could be used to specify them as BH systems. By adding a second color and a third dimension with CCI diagrams separation between systems that contain BHs and systems that contain NSs can be achieved without difficult spectral fitting.
\par
These results suggests a universality in the physical processes driving sources to their location in 3D CCI diagrams. The luminosity of XRBs depends on the mass accretion rate onto the compact object, the strength of the magnetic field, and the mode of accretion. 
\par
In Section 5, we demonstrated that CCI diagrams can be used to distinguish between systems containing NS or BH. In other work, we expand on \cite{gopalan2015} and use several machine-learning techniques to determine the probability that our classifications using CCI are correct (de Beurs et al. ApJ submitted), and use CCI diagrams to classify the nature of ULXs detected in {\it Chandra} observations of external galaxies (Vrtilek et al, in prep)

\section*{Acknowledgements}
We thank the anonymous referee for helpful comments and suggestions. 
 This research has made use of MAXI data provided by RIKEN, JAXA and the MAXI team. This work was supported by the Chandra GO grants (AR5-16007X), Smithsonian 2018 Scholarly Study Program and NASA contract NAS8-03060 (CXC).

\section*{References}
\bibliographystyle{model2-names.bst}\biboptions{authoryear}

\end{document}